\documentclass[aps,pra,twocolumn,amsmath,amssymb,groupedaddress,10pt]{revtex4-1}
\usepackage{tgtermes}

\usepackage{float}
\usepackage{graphicx}  
\usepackage{xcolor}
\usepackage{dcolumn}   
\usepackage{bm}        
\usepackage{amssymb}   
\usepackage{epstopdf}
\usepackage{xfrac}
\usepackage{MnSymbol,wasysym}

\usepackage{adjustbox}

\usepackage{amsmath}
\usepackage{amsfonts}
\definecolor{darkblue}{rgb}{0.1,0.2,0.6}
\definecolor{darkred}{rgb}{0.8,0.1,0.2}
\usepackage[colorlinks,citecolor=darkblue,linkcolor=darkred,urlcolor=darkblue]{hyperref} 
\usepackage{tikz}
\usetikzlibrary{patterns}
\usetikzlibrary{calc,positioning}
\usepackage{enumerate}
\usepackage{setspace}
\usepackage{url}  
\usepackage{mathtools}

\begin{document}

\title{Dynamical renormalization group approach to the spin-boson model}
\author{Hassan Shapourian}
\affiliation{Department of Physics, University of Illinois at Urbana-Champaign, Urbana Illinois 61801, USA}
\date{\today}

\begin{abstract}
We develop a semi-analytical approach beyond the Born-Markov approximation to study the quench dynamics of the spin-boson model in the strong-coupling regime ($\alpha\leq1/2$) for the Ohmic bath.
The basic idea in our approach is to write an effective time-dependent model for the dynamics of the system coupled to the bosonic bath after integrating out high-frequency bath modes. By applying this procedure to the Heisenberg equations of motion, we derive a set of flow equations for the system parameters as a function of time.
The final flow equations look similar to those of the equilibrium renormalization-group theory; however, in our derivation the scaling parameter is set by the real time.
We solve the equations of motion with time-dependent renormalized parameters and show that the resulting dynamics is in decent agreement with the exact NRG calculations as well as the non-interacting blip approximation that is a well-known good solution in this limit.
\end{abstract}

\maketitle

\section{\label{} Introduction}
The decoherence and relaxation processes in quantum mechanics have been introduced to describe the non-Hamiltonian dynamics of open quantum systems due to their coupling to the surrounding environment~\cite{Leggett1987,Weiss1999,Breuer2007}. Typically, the rates associated with these processes, i.e., the relaxation rate and the decoherence (dephasing) rate, are calculated through the standard perturbative expansions in terms of the coupling coefficient to the environment (bath). An archetypal example of the open quantum systems is the spin-boson model which is used as an approximate model to describe a wide range of quantum dissipative phenomena from defect tunneling in solids to electron transfer in chemical and biological reactions~\cite{Grabert,PhysRevLett.68.998,Chandlerbook,Maybook,Nitzanbook,Engel2007,Calhoun2009,Collini2009,ColliniScience2009,Collini2010,Panitchayangkoon2010}. 
In particular, one interest in the spin-boson model is in the context of quantum optics where this model is used to describe
the spontaneous emission of a two-level system (qubit) to its surrounding electromagnetic modes, which can be realized in a variety of systems using real or artificial atoms.
Historically, the perturbative method of the Born-Markov approximation is known to be quite satisfactory in calculating the decay rates in the weak-coupling regime.
However, the emergence of new experimental realizations using artificial structures~\cite{Goldhaber1998,Devoret2002} with high degrees of controllability has brought an access to unusual regimes of spontaneous emission with strong coupling to the bath~\cite{Bourassa2009,Sundaresan}. In particular, there are several proposals for realization of such systems in the superconducting circuits (circuit-QED)~\cite{LeHur2012,Goldstein2013,Peropadre}. 
In addition, the versatile optical manipulation available in these experiments offer new opportunities to design novel procedures at which one can tune a parameter suddenly (quench dynamics)~\cite{Tureci2011,Latta2011}, or drive the system far from the equilibrium~\cite{Sbierski2012}, and hence allow one to study the non-equilibrium phenomena in strongly correlated photonic systems~\cite{Koch_rev,LeHur_Marco}.
In case of the spin-boson model in the strong-coupling regime which is the focus of this work, a typical circuit-QED setup would consist of a qubit coupled to a long cavity. Remarkably, such a setup has been successfully realized very recently~\cite{Sundaresan}.

From a theoretical point of view, non-equilibrium dynamics in the strong-coupling regime has regained much attention recently. The progress in this field is facing two major challenges: first, the non-equilibrium physics cannot be addressed via extensively studied equilibrium formalism and requires a fundamentally more general and possibly more sophisticated approach; second, perturbative methods like the Born-Markov approximation fail in this regime and non-perturbative methods have to be developed. 
The spin-boson model, as a prototypical model for the study of non-equilibrium dynamics, consists of a two-level system (TLS) that is linearly coupled to an infinitely large bosonic bath, where the Hamiltonian is given by $H_{SB} = H_s+ H_b + V$, 
\begin{align} \label{eq:SBmodel}
H_{SB} =- \frac{\Delta_0}{2} \sigma_x + \sum_k \omega_k b_k^\dagger b_k+ \frac{\sigma_z}{2} \sum_k g_k (b_k+b_k^\dagger) 
\end{align}
where $H_s$ and $H_b$ denote the Hamiltonian of the TLS and the bath, respectively, $V$ is the system-bath coupling term, $\Delta_0$ is the TLS frequency, $\sigma_z=|e\rangle\langle e| - |g\rangle\langle g|$ and $\sigma_x= \sigma_+ + \sigma_-= |e\rangle\langle g| + |g\rangle\langle e|$ are TLS operators in which $|e\rangle$ and $|g\rangle$ show the excited state and ground state of TLS respectively. The operators of $k$-th mode of the bosonic bath with frequency $\omega_k$ are denoted by $b_k$ and $b_k^\dagger$ and are linearly coupled to TLS by the coupling coefficient $g_k$. The spin-boson model is characterized by the spectral function
\begin{align} \label{eq:spec}
J(\omega)= \pi \sum_k g_k^2 \delta (\omega-\omega_k) = 2\pi \alpha \omega_c^{1-s} \omega^s e^{-\omega/\omega_c}   
\end{align}
where $\alpha$ is a dimensionless coupling constant, $s$ is a real positive number and bath frequencies are bounded by a smooth cut-off frequency $\omega_c$. We assume the linear dispersion $\omega_k=c|k|$ where we set $c=1$. Our focus in this work is the Ohmic bath where $s=1$. It is worth mentioning that the continuous spectral function is not directly related to the experimental setup~\cite{Sundaresan} cited earlier and one way to realize a continuous spectral function is possibly when the frequency spacing between successive modes of the long cavity is smaller than the linewidth of each mode. 

The spin-boson model in the strong-coupling regime has been extensively studied over so many years and except for some special parameter values an exact solution is not known. 
In an original attempt by Leggett {\it et al}.~\cite{Leggett1987}, they developed a method  known as the non-interacting blip approximation (NIBA), where a power-series expansion in $\Delta_0^2$ is carried out which gives very reliable results for the diagonal matrix elements of the reduced density matrix of TLS over a certain range of the coupling constant $\alpha<1/2$.
Since then, there have been several successful computational methods to address this problem.
The quantum Monte Carlo calculations~\cite{Egger1992,Egger_QMC,Stockburger_QMC} have provided reliable results for the full reduced density matrix.
There have also been various techniques based on the mapping of the Ohmic spin-boson model onto the anisotropic Kondo model where one can solve the latter exactly using the numerical renormalization group (NRG)~\cite{Costi_NRG,*Costi_NRG2,Costi_NRG3}, Bethe ansatz~\cite{Costi_Bethe}, or the conformal field theory~\cite{Lesage_CFT}. Another type of NRG method have also been devised specially for the bosonic baths~\cite{Bulla2003,Bulla2005,Anders2006, Anders2007,LeHur2011,Guo2012}. The method of the real-time renormalization-group~\cite{Keil2001,Kashuba2013} has been shown to be quite effective in calculating the dynamics of the TLS full density matrix.
A non-perturbative stochastic method~\cite{Orth2013,Orth2010,Henriet2014} as a numerically exact method have also been developed recently. This method is quite general and it has been extended to compute various correlation functions in the presence of an external drive. 

Another important effort along solving the dynamics of the spin-boson model is known as the flow equation approach~\cite{Kehrein_1997,Kehrein_1998,Kehrein2008} that is formulated in terms of a sequence of infinitesimal unitary transformations which yields a set of flow equations for the model parameters as seen in the renormalization-group (RG) methods.
Motivated by the observation that the bare model parameters flow as a function of time, we would like to present a complementary derivation of this flow, which we call (real-time) dynamical renormalization-group approach. 
Throughout this method, we assume that the dynamics of TLS can be given in terms of a time-local master equation in which the TLS frequency and decay rate are slowly varying  in time. In the process of computing the TLS parameters, we show that this assumption is self-consistent for the spin-boson model. 
Our calculation is based on the idea that fast bath modes with high frequencies can be averaged to yield an effective correction for the TLS dynamics. As time evolves, more bath modes are averaged out and at sufficiently long time the TLS parameters reach their steady-state values which are very close to effective TLS frequency and decay rate given by NIBA. In other words, the flow equations provide a dynamical transition from the bare values to renormalized values of the TLS parameters. 
This method is designed for the range of coupling constant up to $\alpha\leq1/2$. 
We compare our results with NIBA as well as numerically exact NRG calculations~\cite{Guo2012} and obtain a remarkable agreement.
We also benchmark our method against the case of spin-boson model after the rotating-wave approximation which can be solved exactly. Interestingly, we observe a good agreement with the exact results in this case too.

In the rest of this paper, we present background and details of our calculations and compare our approach with other known almost exact results in Sec.~\ref{sec:method} and draw conclusions and propose possible applications and extensions of our method in Sec.~\ref{sec:conc}.

\section{\label{sec:method} Method and Results}

\begin{figure}
\includegraphics[scale=1]{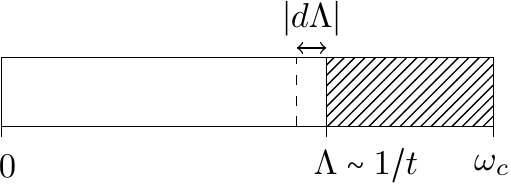}
\caption{\label{fig:bathspec} At time $t$, the spectrum of bath modes is divided into a high-frequency part $\omega>\Lambda$ (hatched region) and a low-frequency part $\omega <\Lambda$. The effect of the high-frequency bands on the system dynamics can be accounted by putting their average value.
At $t+dt$, a new slice of low-frequency sector (shown by a dashed line) has to be removed from the low-frequency sector and added as a high-frequency contribution.} 
\end{figure}

In this section, we explain our method and show that it gives consistent results which are quantitatively in good agreement with the NIBA method as well as the NRG results.
Let us begin with a formal statement of the problem and a proposed form of the solution which motivates this work.
We would like to study the quench dynamics of the spin-boson model with an initially factorizing state in which the full density matrix is a tensor product $\rho(t=0)= \rho_s (0)\otimes \rho_b(0)$. The goal is to compute the reduced density matrix of TLS $\rho_s(t)=\text{tr}_b(\rho(t))$ for $t>0$ where $\text{tr}_b(\ )$ means to trace out the bath degrees of freedom. 
The usual method to find $\rho_s(t)$ is to start with the Von Neumann equation $\partial_t \rho= -i[H_I (t),\rho]$ for the full density matrix in the interaction picture where $H_I(t)= e^{i H_0} V e^{-i H_0}$ and $H_0=H_s+H_b$ defined in Eq.~(\ref{eq:SBmodel}). After a straightforward manipulation, one arrives at
\begin{align} \label{eq:master}
\frac{\partial \rho_s}{\partial t} = - \int_0^t dt'\  \text{tr}_b\left( [H_I(t),[H_I(t'),\rho(t')]] \right).
\end{align}
The above expression is exact up to assuming that $\text{tr}_b ([H_I,\rho(0)])=0$ which is always the case for a thermal bath. From this point, various approximations are made to solve for $\rho_s(t)$. The Born approximation is
to write $\rho(t')\approx \rho_s(t')\otimes \rho_b(0)$ in the integrand. The Redfield approximation builds on top of the Born approximation by writing $\rho(t')\approx \rho_s(t)\otimes \rho_b(0)$ and removing the density matrix from the integrand.
Finally, the Born-Markov approximation goes one step further by assuming that the result of the integral is $t$ independent and $H_I(t')$ should be replaced by $H_I(t-t')$. We note that the Redfield and Markov approximations look similar in the sense that both equations are local in time, i.e. only $\rho_s(t)$ is present on the right-hand side of the equation, while the important difference between them is that the former involves time-dependent coefficients while all coefficients are time independent in the latter.

Inspired by these approximations, here we propose that the dynamics of the system can be effectively described by a time-local master equation which is characterized by  the effective time-dependent TLS frequency  $\Delta(t)$ and decay rate $\gamma(t)$ to be determined shortly. 
Interestingly, the exact solution of the spin-boson model after the rotating-wave approximation (also called the spontaneous emission) can be written in this form~\cite{Breuer2007,Hu2000}. 
Here we present a method to systematically derive approximate solutions for $\Delta(t)$ and $\gamma(t)$ through an RG-type argument based on coarse graining over the real time.
There is also a general method known as the time-convolutionless projection operator technique which has been developed in \cite{Shibata1977,*Shibata1979,*Shibata1980} to derive the time-local equations for the dynamics of system. This method has been successfully applied to various models and has been shown to capture dynamics beyond the Born-Markov approximation (see also~\cite{Breuer2007}). 


Let us now derive the defining equations for the system parameters $\Delta(t)$ and $\gamma(t)$. 
To this end, it is more convenient to work with the Heisenberg equations of motion instead of the density matrix dynamics in Eq.~(\ref{eq:master}). 
The ultimate form of equations of motion for the TLS operators is
\begin{eqnarray}
\label{eq:sigmaxdotf} \partial_t \langle{\sigma}_x \rangle&=&- \gamma(t) (\langle\sigma_x\rangle- 1),\\
\label{eq:sigmaydotf} \partial_t \langle{\sigma}_y \rangle&=& \Delta(t) \langle\sigma_z \rangle- \gamma(t) \langle\sigma_y \rangle, \\
\label{eq:sigmazdotf} \partial_t \langle{\sigma}_z \rangle&=& -\Delta(t) \langle\sigma_y \rangle,
\end{eqnarray}
where $\langle ...\rangle=\text{tr}(\rho_s(t) ...)$ is the expectation value of a system operator, $\Delta(t)$ and $\gamma(t)$ are given in terms of the flow equations. These equations can be effectively interpreted as the TLS being dressed up with the high-frequency  bath modes as the time evolves. In this respect, we call this method the dynamical renormalization group (DRG). The current scheme is inspired by the idea developed in~\cite{Ludwig2010} regarding the dynamics of vortices in 2D sine-Gordon model where the equations of motion correspond to classical fields.
The idea is that at time $t$ the bath modes are divided into the low-frequency $\omega t \lesssim 1$ and the high-frequency $\omega t\gtrsim 1$ sectors and their effect on the dynamics of TLS are treated separately (Fig.~\ref{fig:bathspec}).
Then, the coarse-graining scheme follows by replacing the high frequency bath modes with their effective average values.

The DRG process starts with the Heisenberg equations of motion
\begin{eqnarray}
\label{eq:sigmaxdot} \partial_t {\sigma}_x&=& -\sum_k g_k ( b_k+  b^{\dag}_k) \sigma_y ,\\
\label{eq:sigmaydot} \partial_t {\sigma}_y &=& \Delta_0 \sigma_z + \sum_k g_k (b_k + b^\dag_k) \sigma_x ,\\
\label{eq:sigmazdot} \partial_t {\sigma}_z &=& -\Delta_0 \sigma_y , \\
\label{eq:bkdot} \partial_t {b}_k &=& -i \omega_k b_k - i g_k \frac{\sigma_z}{2}.
\end{eqnarray}
It is evident that this set of equations cannot be closed and every time we replace one equation into another, higher-order terms (so-called secular terms) are generated due to the nonlinearity of the spin operators. The usual perturbation theory would be to cut this process at some step and decouple TLS operators from the bath operators; for instance, keeping terms up to the second order in $g_k$ yields the Born approximation.  However, this approximation clearly fails in the strong-coupling regime. 
It is worth noting that in contrast to the original basis of the Hamiltonian in Eq.~(\ref{eq:SBmodel}), in the displaced bath oscillators basis, decoupling the spin and bath operators in the Heisenberg equations of motion yields the same results as NIBA~\cite{Dekker1987}. 
As mentioned earlier, our objective here is to use the original basis of the spin-boson Hamiltonian and to keep the form of the equations the same as that of the Born-Markov approximation, Eqs.~(\ref{eq:sigmaxdotf})-(\ref{eq:sigmazdotf}), but modify the parameters accordingly as time evolves. 

We now apply the following renormalization procedure to the equations of motion. We rescale the time as $t\to t(1+d\Lambda/\Lambda)$ and the bath mode frequencies as $k \to k(1-d\Lambda/\Lambda)$. This implies that the bath cutoff frequency is rescaled as  $\Lambda \to \Lambda' \equiv \Lambda(1-d\Lambda/\Lambda)$, so the bath modes between $\Lambda'$ and $\Lambda$ are to be removed. 
The removal process of high-frequency bath modes is done by finding the effective term generated by these modes after averaging them through a perturbative treatment. As expected, the newly generated terms after one RG step is nothing but a correction to the TLS frequency and decay rate.
In order to do this, we substitute the solutions of $b_k$ and $b_k^\dagger$ for $\Lambda'<\omega_k< \Lambda$ in the equations of motion for TLS, and we keep only terms up to the second order in terms of the coupling $g_k$. As a result of time averaging and after taking the expectation value over the bath modes, the high-frequency modes  contribute as a Lamb shift correction to TLS frequency as well as a Markovian decay rate (see Appendix A for detailed derivations). In other words, as we go to an infinitesimally later time $t(1+d\Lambda/\Lambda)$, a narrow band of modes (see Fig.~\ref{fig:bathspec}) in the low-frequency sector must be transferred to the high-frequency sector and this modifies ${\Delta}$ to ${\Delta}+d\Delta$ and ${\gamma}$ to ${\gamma}+d\gamma$. Consequently, one can calculate the flow equations as
\begin{align}
\frac{d\Delta}{d\Lambda} =& \frac{\Delta}{2\pi} \frac{J(\Lambda) (\Delta^2-\Lambda^2)}{(\Delta^2-\Lambda^2)^2+ \gamma^2\Lambda^2}, \\
\frac{d\gamma}{d\Lambda} =& \frac{\Delta^2}{\pi} \frac{\gamma J(\Lambda) }{(\Delta^2-\Lambda^2)^2+ \gamma^2\Lambda^2}.
\end{align}
Remarkably, the flow equation of $\Delta$ when $\gamma\to 0$ coincides with the equilibrium perturbative RG equations of spin-boson model which yields the Kondo energy scale $T_K=\Delta_0 (\Delta_0/\omega_c)^{\alpha/(\alpha-1)}$ for TLS in the scaling limit where $\Delta_0 \ll \omega_c$~\cite{Leggett1987,Weiss1999}.

\begin{figure}
\includegraphics[scale=0.85]{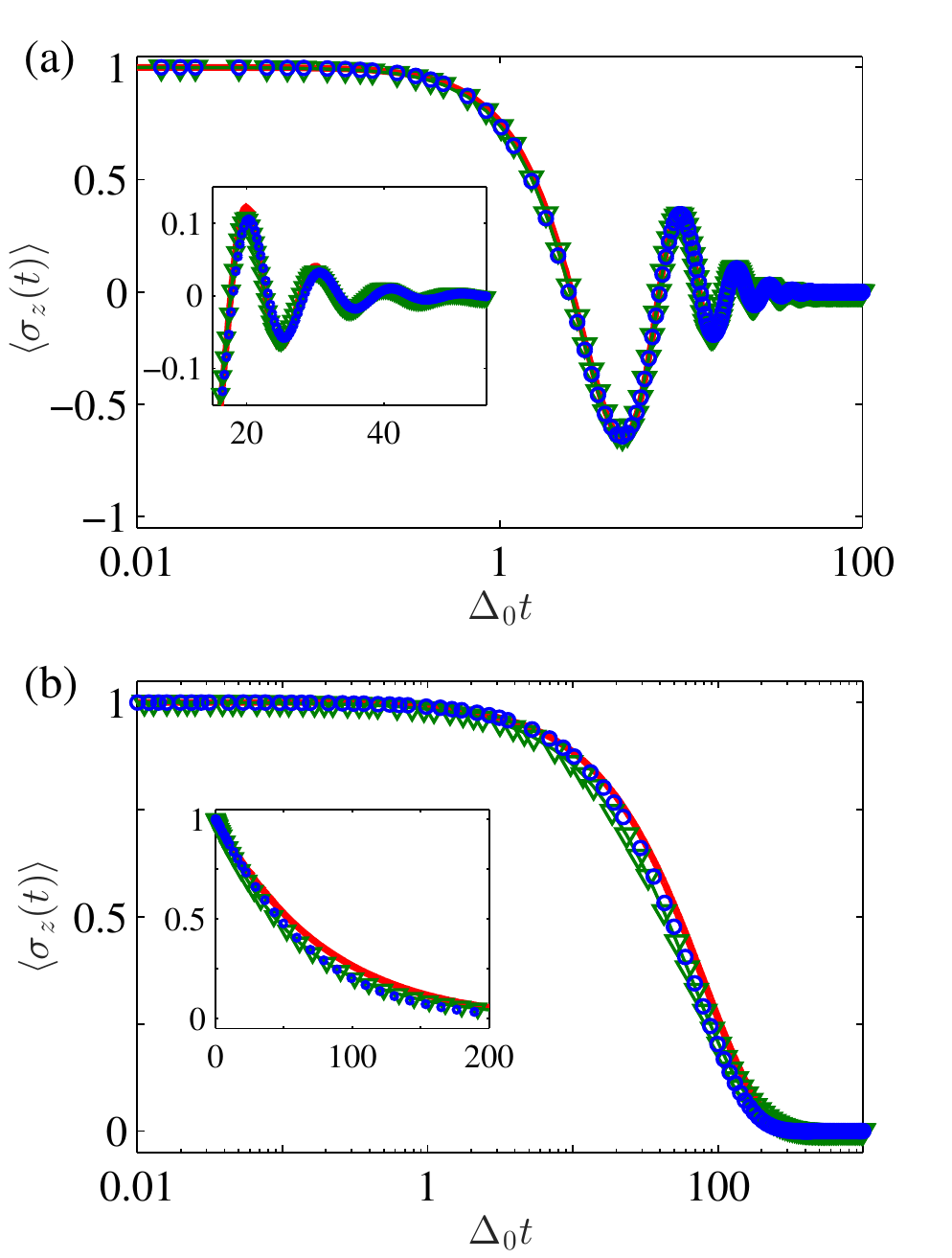}
\caption{\label{fig:compSB}  (Color online) Comparison of TLS dynamics  $\langle\sigma_z(t)\rangle$
 for (a) $\alpha=0.10$  and (b) $\alpha=0.5$ (the Toulouse point). The inset shows the linear time scale. In all figures, blue circles are DRG, green triangles are NIBA and solid red is NRG. The NRG results are courtesy of Weichselbaum (see Ref.~\cite{Guo2012}). Note that NIBA is the exact analytical solution at the Toulouse point. $\eta$ is set to (a) $1.0$ and (b) $0.5$. Here, $\Delta_0/ \omega_c= 0.01$.}
\end{figure} 

\begin{figure}
\includegraphics[scale=0.85]{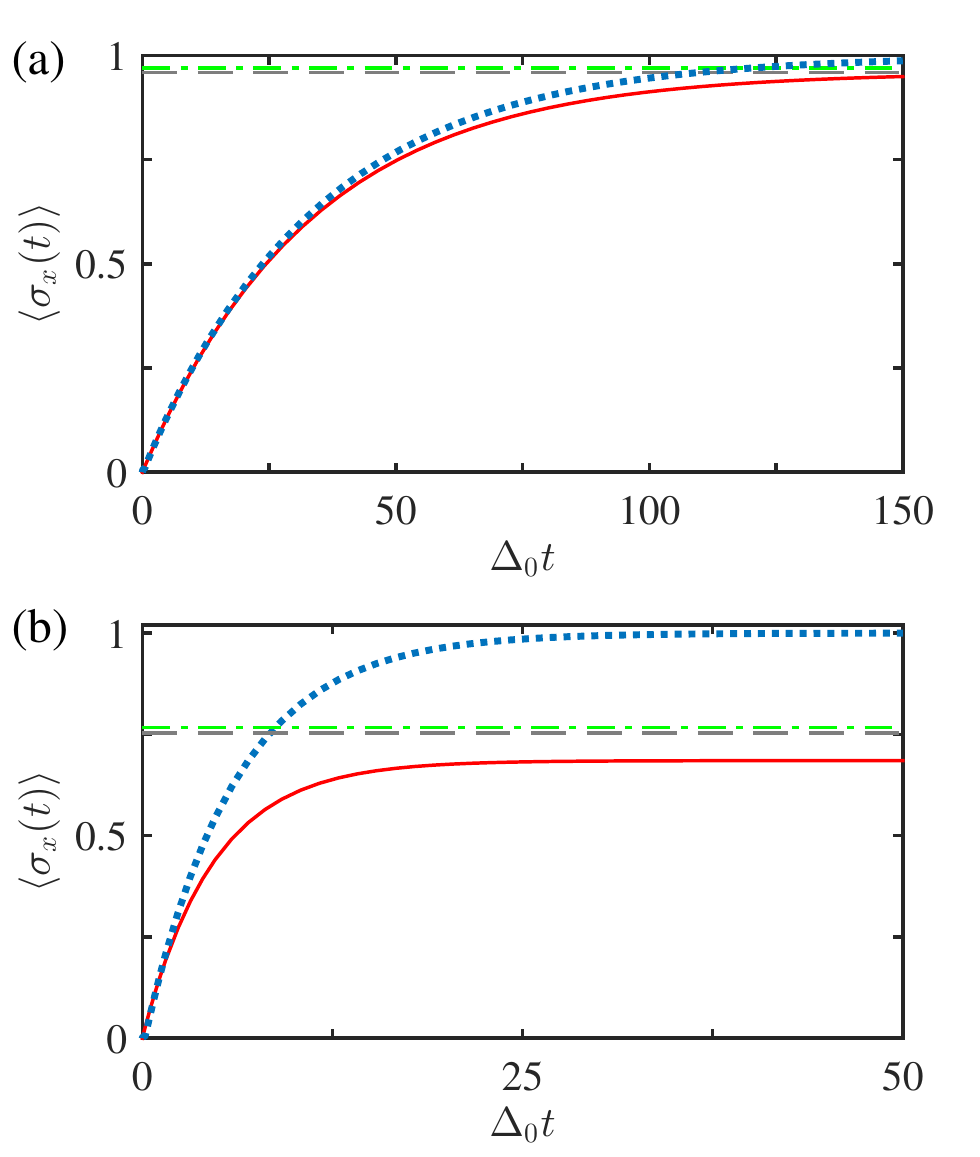}
\caption{\label{fig:sx_compSB}  (Color online) Comparison of TLS dynamics $\langle\sigma_x(t)\rangle$
 for (a) $\alpha=0.01$  and (b) $\alpha=0.1$. In both figures, the dotted blue is DRG and the solid red is the weak-coupling theory beyond the NIBA~\cite{Weiss1999,Weiss1989_1,Weiss1989_2}. The long-time thermodynamic value $\langle\sigma_x (\infty)\rangle$ predicted by the Bethe ansatz~\cite{LeHur_Bethe} (dashed gray) and a rigorous Born approximation~\cite{DiVincenzo} (dash-dotted green) are shown for reference. Here, $\eta$ is set to $1.0$ in both cases and $\Delta_0/ \omega_c= 0.02$.}
\end{figure}

\begin{figure}
\includegraphics[scale=0.85]{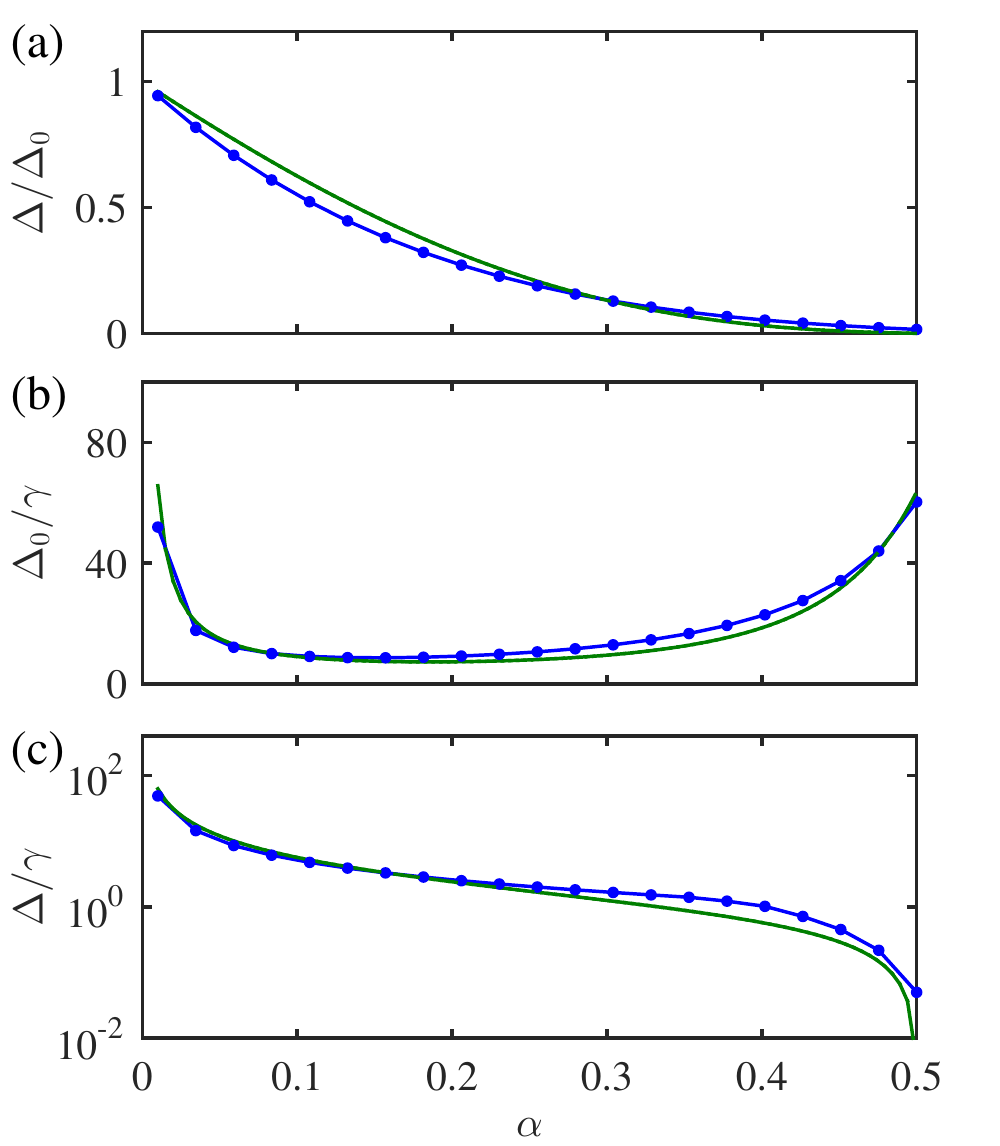}
\caption{\label{fig:longT_comp}  (Color online) Comparison of the long-time values for (a) the renormalized TLS frequency $\Delta$ , (b) relaxation time $\tau=\gamma^{-1}$, and (c) the quality factor using DRG (blue circles) and NIBA (solid green).}
\end{figure} 

Next, the time-dependent parameters are used to evaluate the dynamics of TLS through Eqs.~(\ref{eq:sigmaxdotf})-(\ref{eq:sigmazdotf}).  
A comparison between different methods is illustrated in Figs.~\ref{fig:compSB} and \ref{fig:sx_compSB}.
Here, the TLS is initially prepared in the excited state $|e\rangle$ and the bath is at the zero temperature.
As a first benchmark, we compare the dynamics of $\sigma_z$ in Fig.~\ref{fig:compSB}.
The NIBA method is known to yield very good (almost exact~\cite{Weiss1999}) results for the dynamics of $\sigma_z$ and the agreement between our DRG and NIBA is quite remarkable.
  Interestingly, for $\alpha=0.5$ (Toulouse point) the agreement with NIBA which is the exact solution at this point (Fig.~\ref{fig:compSB}(b)) makes the DRG method even more promising.
 As a second benchmark, we compute the dynamics of $\sigma_x$ in Fig.~\ref{fig:sx_compSB} where a comparison with the weak-coupling theory beyond the NIBA~\cite{Weiss1999,Weiss1989_1,Weiss1989_2} and the thermodynamic expectation value $\langle\sigma_x(\infty)\rangle$ at long times, using Bethe ansatz~\cite{LeHur_Bethe} and a rigorous Born approximation~\cite{DiVincenzo}, are shown for reference. It is evident from Eq.~(\ref{eq:sigmaxdotf}) as well as Fig.~\ref{fig:sx_compSB} that the long-time value of DRG is always $1$ which is different from the prediction of Bethe ansatz~\cite{LeHur_Bethe}. This probably indicates that the DRG method at this order of approximation (single-loop RG) is best for evaluating $\sigma_z$ and not $\sigma_x$. Nevertheless, it is worth noting that DRG does not have the two well-known issues of the NIBA in the long-time regime of the unbiased spin-boson model; namely, (i) the long-time value $\langle\sigma_x(\infty)\rangle$ is not singular; (ii) there is no power-law behavior for the long-time $\langle\sigma_z(t)\rangle$ and DRG simply converges to a damped oscillator as anticipated from the conformal field theory~\cite{Lesage_CFT}.

We should note that in our arguments above the real time is related to the cutoff frequency $\Lambda$ up to a proportionality coefficient $\eta$, i.e., $\Lambda = \eta/ t$. This time-independent coefficient determines how fast the integration process needs to be done; i.e., the smaller the $\eta$, the faster the spectrum is swept and the parameters flow away from short-time solutions towards the long-time renormalized values more quickly. Interestingly, the long-time values of effective parameters $\Delta$ and $\gamma$ do not depend on the choice of $\eta$. However, to make the DRG results match the transient dynamics, $\eta$ must be optimized for any given spectral function $J(\omega)$. In this sense, $\eta$ can be regarded as a variational parameter $0<\eta\leq1$. We obeserve that the general trend is as follows:  $\eta$ must be decreased as $\alpha$ is increased and it does not depend on $\Delta_0/\omega_c$.
Explicit examples for $\eta$ dependence are shown in Fig.~\ref{fig:eta} of Appendix~\ref{sec:DRG_der}.
 Figure~\ref{fig:longT_comp} shows the long-time renormalized TLS frequency $\Delta$ and relaxation time $\tau=\gamma^{-1}$ as a function of the coupling coefficient $\alpha$, where there is an acceptable agreement between the DRG and NIBA results. Furthermore, as we see in Fig.~\ref{fig:longT_comp}(c), the quality factor remains rather close to that of NIBA, which is in turn confirmed by the conformal field theory~\cite{Lesage_CFT}.

Few remarks about the DRG method compared to the other recently developed methods are in order.
The short-time dynamics is explicitly built into the DRG steps and the long-time dynamics is shown to match with the NIBA results. These two properties make the DRG method a good approximation for both short-time transient dynamics as well as the long-time dynamics. This must be contrasted from other interesting variational approaches~\cite{Nazir2012,Zhao2013,Bera2014} where a unitary transformation is carried out and a perturbative expansion is performed in the rotated basis to describe the quench dynamics. As also pointed out in Refs.~\cite{Nazir2012,Zhao2013}, this expansion is not guaranteed to give a reliable short-time dynamics.
Moreover, in the variational methods it is usually assumed that the long-time interaction with bath relaxes the final state of the global system (TLS+bath) to the ground state of the full Hamiltonian (also called shifted thermal bath). As a result of this assumption, it is relevant to expand the interactions perturbatively around a good variational ansatz for the ground state. Interesting enough, we do not make such an assumption in the DRG method and this long-time behavior naturally comes out of the flow equations.  It should be noted that the DRG method shares some similarities to the flow equation approach~\cite{Kehrein2008} mentioned earlier and we believe that our approach is quite comparable with this method.
 
To further benchmark our DRG method, we have also applied it to the spontaneous emission model which can be solved exactly and have compared the DRG results with the exact solution where we have found an excellent agreement (see Appendix~\ref{sec:SE}). 

\section{\label{sec:conc} Discussion}

In conclusion, we present a semi-analytical approach to study the dynamics of the spin-boson model in the strong-coupling regime. The key idea is to represent the dynamics of TLS in terms of a time-local master equation with the time-dependent frequency and decay rate. In other words, the TLS dynamics can be described by an effective time-dependent model in which the TLS parameters are determined by a set of flow equations where the real time is the flow parameter. The flow equations are derived by doing partial averaging over fast oscillating high-frequency bath modes. We observe that the long-time values of TLS frequency and decay rate match with the renormalized values computed by the NIBA method over a wide range of coupling constant $0<\alpha\leq1/2$.
In terms of the quench dynamics of TLS, the DRG approach has demonstrated a remarkable agreement with the other reliable methods well beyond the perturbative Born-Markov approximation. 

As a possible application of the DRG approach, it can be generalized to more complex out-of-equilibrium situations where the system is strongly driven by an external coherent source. This could involve calculating the correlation functions beyond the perturbative expansions. Another possible interesting direction is the generalization to the dynamics of a biased TLS, including terms like $\frac{\epsilon}{2} \sigma_z$.
We should note that the flow equation in our method fails for larger coupling constants $\alpha>1/2$, which is also the case in several other approaches. The extension of the DRG method to the ultrastrong-coupling regime is an interesting direction to pursue.

In principle, the general ideas and treatments discussed in the DRG method can be applied to other impurity systems such as  the quench dynamics of the Anderson impurity model with the rich phase diagram consisting of three fixed points: the free orbital, local moment, and strong coupling. One interesting possibility is to investigate the dynamical crossovers from one fixed point to another~\cite{Tureci2009}.

\acknowledgements
I would like to acknowledge insightful discussions with H. E. T\"ureci, L. Mathey, M. Schir\'o, K. Le Hur, and A. Polkovnikov.
I also thank B. Bruognolo and A. Weichselbaum for sharing the numerical data which are used in Fig.~\ref{fig:compSB}.


\newpage

\begin{appendix}

\begin{widetext}

\section{\label{sec:DRG_der} Derivation of the flow equations for the spin-boson model}
In this appendix we give a detailed derivation of the quench dynamics in the spin-boson model using the DRG method.
Following the idea introduced in~\cite{Ludwig2010} (originally for the dynamics of classical observables), we would like to derive a set of flow equations via the Heisenberg equations of motion. 
The process of removing high-frequency bath modes begins with substituting the solutions of $b_k$ and $b_k^\dagger$ for $\Lambda'<\omega_k< \Lambda$ in the equations of motion for TLS. The full solution of $b_k$ can be easily found to be
\begin{equation}\label{eq:bkorig}
b_k(t)= b_k(0) e^{-i\omega_k t}-i\frac{g_k}{2} \int_0^t dt' e^{-i \omega_k (t-t')} \sigma_z(t').
\end{equation}
We are going to treat the high-frequency shell perturbatively and show that these modes ultimately lead to corrections in the TLS dynamics. In other words, we split the bath modes as in
\begin{eqnarray}
\label{eq:sigmaxdot} \partial_t {\sigma}_x&=& -\sum_{k<\Lambda'} g_k ( b_k+  b^{\dag}_k) \sigma_y
-\sum_{k\in d\Lambda} g_k ( b_k+  b^{\dag}_k) \sigma_y ,\\
\label{eq:sigmaydot} \partial_t {\sigma}_y &=& \Delta_0 \sigma_z + \sum_{k<\Lambda'} g_k (b_k + b^\dag_k) \sigma_x 
+ \sum_{k\in d\Lambda} g_k (b_k + b^\dag_k) \sigma_x ,\\
\label{eq:sigmazdot} \partial_t {\sigma}_z &=& -\Delta_0 \sigma_y.
\end{eqnarray}
where $k \in d\Lambda$ refers to the fast oscillating shell of bath modes $\Lambda'<k<\Lambda$.
Plugging Eq.~(\ref{eq:bkorig}) into the high-frequency part of the right-hand side (RHS) of Eqs.~(\ref{eq:sigmaxdot}) and (\ref{eq:sigmaydot}) yields
\begin{align}
\label{eq:sigmaxRG1}
f^>_x \equiv & -\frac{1}{2} \sum_{k \in d\Lambda} g_k {\Big[} ( b_k(0) e^{-i\omega_k t}+  b^{\dag}_k(0) e^{i\omega_k t} ) \sigma_y(t) +  \sigma_y(t) ( b_k(0) e^{-i\omega_k t}+  b^{\dag}_k(0) e^{i\omega_k t} ) {\Big]}  \nonumber \\
&+ \frac{1}{2} \sum_{k \in d\Lambda} g_k^2 \int_0^t \text{d}t' \text{sin}(\omega_k(t-t')) \ (\sigma_z(t')\sigma_y(t) + \sigma_y(t)\sigma_z(t')), \\
\label{eq:sigmayRG1}
f^>_y \equiv & \frac{1}{2} \sum_{k \in d\Lambda} g_k {\Big[} ( b_k(0) e^{-i\omega_k t}+  b^{\dag}_k(0) e^{i\omega_k t} ) \sigma_x(t) +  \sigma_x(t) ( b_k(0) e^{-i\omega_k t}+  b^{\dag}_k(0) e^{i\omega_k t} ) {\Big]}  \nonumber \\
&- \frac{1}{2} \sum_{k \in d\Lambda} g_k^2 \int_0^t \text{d}t' \text{sin}(\omega_k(t-t')) \ (\sigma_z(t')\sigma_x(t) + \sigma_x(t)\sigma_z(t')), 
\end{align}
Note that we wrote the equations in a symmetric form, since the equal-time bath operators commute with TLS operators in the original Eqs.~(\ref{eq:sigmaxdot})-(\ref{eq:sigmazdot}). 
We first show the detailed derivation of the renormalized parameters for $\sigma_y$, the same derivation can be done for $\sigma_x$. Later, we will discuss the additional term $\gamma(t)$ on the RHS of Eq.~(\ref{eq:sigmaxdotf}) which only appears in the equation of motion for $\sigma_x$. 

The next step to find the effective dynamics of $\sigma_y$ is to substitute the $\sigma_x$ dynamics in Eq.~(\ref{eq:sigmayRG1}) and keep only terms up to the second order of $g_k$. 
The free dynamics of $\sigma_x$ to the leading order in $g_k$ is given by
 \begin{align}
\label{eq:sigmaxpert}\sigma_x(t)= e^{-\gamma_x t} \sigma_x(0) 
- \sum_{k} g_k \int_0^t \text{d}t' e^{-\gamma_x (t-t')} (b_k(0)e^{-i\omega_k t'}+ b_k^\dagger(0)e^{-i\omega_k t'}) \sigma_{y} (t')+O(g_k^2).
\end{align}
We substitute this into Eq.~(\ref{eq:sigmayRG1}),
\begin{align} \label{eq:RGstep2}
f^>_y=&- \frac{1}{2} \sum_{k',k \in d\Lambda} g_k g_{k'} \int_0^t \text{d}t' \ e^{-\gamma_x (t-t')} {\Big[} ( b_k(0) e^{-i\omega_k t}+  b^{\dag}_k(0) e^{i\omega_k t} )( b_{k'}(0) e^{-i\omega_{k'} t'}+  b^{\dag}_{k'}(0) e^{i\omega_{k'} t'} ) \sigma_y(t') \nonumber \\ &+  \sigma_y(t') ( b_{k'}(0) e^{-i\omega_{k'} t'}+  b^{\dag}_{k'}(0) e^{i\omega_{k'} t'} ) ( b_k(0) e^{-i\omega_k t}+  b^{\dag}_k(0) e^{i\omega_k t} ) {\Big]}.
\end{align} 
We are interested in corrections up to the second order with respect to the coupling coefficients $g_k$; hence, the $\sigma$ operators in the second line of Eq.~(\ref{eq:sigmayRG1}) are to be replaced by the zeroth order solution given by the free (decoupled) dynamics. In fact, these terms vanish at this order of approximation due to the anti-commuting property of Pauli matrices; \emph{i.e.} $\{\sigma_i,\sigma_j\}=0$ for $i \neq j$. We should note that this, however, is not the case for $f^>_x$ (see Eq.~(\ref{eq:intx2})) and indeed it is this non-zero value which gives rise to the term $\gamma(t)$ in Eq.~(\ref{eq:sigmaxdotf}).
After taking the expectation value over the bath modes in Eq.~(\ref{eq:RGstep2}), we arrive at
\begin{align} \label{eq:RGstep3}
\langle f^>_y\rangle = - \sum_{k\in d\Lambda} g_k^2 \ (2 n_b(\omega_k)+1) \int_0^t \text{d}t' e^{-\gamma_x (t-t')}\text{cos}(\omega_k(t-t')) \langle\sigma_y(t')\rangle.
\end{align}
Here, we assume that the bath is in the thermal equilibrium with temperature $T$ and $n_b(\omega_k) = \langle b_k^\dagger b_k \rangle=1/[\exp({\hbar \omega_k}/k_B T)-1]$.
Next, we plug the free dynamics of $\sigma_y$ into the above equation, that is
\begin{align} \label{eq:freesigmay}
\sigma_{y}(t') =& e^{-\gamma_y (t'-t)/2} {\Big[} \sigma_y(t) {\big(} \text{cos}(\Omega(t'-t)) - \frac{\gamma_y}{2\Omega} \text{sin}(\Omega(t'-t)) {\big)} +\sigma_z(t) \frac{\Delta}{\Omega} \text{sin}(\Omega(t'-t)) {\Big]} + O(g_k) 
\end{align}
where $\Omega=\sqrt{\Delta^2-\gamma_y^2/4}$. The decay rates $\gamma_x$ and $\gamma_y$ are introduced to make the flow equations self-contained as in any standard RG analyses. It is important to note that the decay rates are equal and we just use different notations for the book keeping. As we will see later, they obey the same flow equations.
As a result of the substitution, we get
\begin{align}
\langle f^>_y\rangle =&  \langle\sigma_z(t)\rangle {\cal F}_z - \langle\sigma_y(t)\rangle {\cal F}_y
\end{align}
where we introduce the following functions
\begin{align}
\label{eq:intz} {\cal F}_z=& \frac{\Delta}{\Omega}  \sum_{k\in d\Lambda} g_k^2 \ (2 n_b(\omega_k)+1)  \int^t \text{d}t' \ e^{-(\gamma_y-2\gamma_x)(t'-t)/2} \text{cos}(\omega(t-t')) \ \text{sin}(\Omega(t'-t)),  \\
\label{eq:inty} {\cal F}_y =& \sum_{k\in d\Lambda} g_k^2 \ (2 n_b(\omega_k)+1) 
\int^t \text{d}t' \ e^{- (\gamma_y-2\gamma_x) (t'-t)/2} \text{cos}(\omega(t-t')) \ {\big(} \text{cos}(\Omega(t'-t)) - \frac{\gamma_y}{2\Omega} \text{sin}(\Omega(t'-t)) {\big)}.
\end{align}
Note that the above functions contain damped oscillating terms which are averaged to zero and the simplified results are
\begin{align}
\label{eq:intz} {{\cal F}_z}=& \frac{\Delta}{2} \sum_{k\in d\Lambda} g_k^2 \ (2 n_b(\omega_k)+1)\frac{\Delta^2-\omega_k^2+\gamma_x(\gamma_x-\gamma_y)}{\Delta^4+(\omega_k^2+\gamma_x^2)(\omega_k^2+(\gamma_x-\gamma_y)^2)-2\Delta^2 (\omega_k^2+ \gamma_x(\gamma_y-\gamma_x))}, \\
\label{eq:inty} {{\cal F}_y} =& \sum_{k\in d\Lambda} g_k^2 \ (2 n_b(\omega_k)+1) \frac{\Delta^2 \gamma_x+ (\omega_k^2+\gamma_x^2) (\gamma_x-\gamma_y)}{\Delta^4+(\omega_k^2+\gamma_x^2)(\omega_k^2+(\gamma_x-\gamma_y)^2)-2\Delta^2 (\omega_k^2+ \gamma_x(\gamma_y-\gamma_x))}.
\end{align}
It is worth noting that the oscillating terms are damped by an exponential factor in Eqs. (\ref{eq:intz}-) and (\ref{eq:inty}) which means that as $t$ becomes large the corrections due to discarding these terms become exponentially small. In addition, for extremely short times $t<1/\omega_c$ the oscillating terms can be neglected, too. This approximation is justified by the fact that the spectral density $J(\omega)\propto e^{-\omega/\omega_c}$ is exponentially small for extremely high frequencies.

Therefore, the two averaged terms add to the self-dynamics of TLS and the corrections are
\begin{align}
d{\Delta} &= \frac{\Delta}{2} \sum_{k\in d\Lambda} \frac{g_k^2 \ (2 n_b(\omega_k)+1) {\Big(}\Delta^2-\omega_k^2+\gamma_x(\gamma_x-\gamma_y){\Big)}}{\Delta^4+(\omega_k^2+\gamma_x^2)(\omega_k^2+(\gamma_x-\gamma_y)^2)-2\Delta^2 (\omega_k^2+ \gamma_x(\gamma_y-\gamma_x))}, \\
\label{eq:longtimegam}
d{\gamma}_y &=   \sum_{k\in d\Lambda} \frac{g_k^2 \ (2 n_b(\omega_k)+1) {\Big(} \Delta^2 \gamma_x+ (\omega_k^2+\gamma_x^2) (\gamma_x-\gamma_y){\Big)}}{\Delta^4+(\omega_k^2+\gamma_x^2)(\omega_k^2+(\gamma_x-\gamma_y)^2)-2\Delta^2 (\omega_k^2+ \gamma_x(\gamma_y-\gamma_x))}.
\end{align}
We focus on the zero temperature case where $n_b(\omega_k)=0$. As we go to an infinitesimally later time $t + dt$, a band of modes in the low frequency sector is added to the high frequency sector (as shown in Fig.~\ref{fig:bathspec}) and this changes ${\Delta}$ to ${\Delta}+d\Delta$ and ${\gamma}_y$ to ${\gamma}_y+d\gamma_y$. Consequently, one can calculate the flow equations as
\begin{align}
\frac{d\Delta}{d\Lambda} &= \frac{\Delta}{2\pi}\frac{J(\Lambda) {\Big(}\Delta^2-\Lambda^2+\gamma_x(\gamma_x-\gamma_y){\Big)}}{\Delta^4+(\Lambda^2+\gamma_x^2)(\Lambda^2+(\gamma_x-\gamma_y)^2)-2\Delta^2 (\Lambda^2+ \gamma_x(\gamma_y-\gamma_x))}, \\
\frac{d\gamma_y}{d\Lambda} &= \frac{1}{\pi} \frac{J(\Lambda) {\Big(} \Delta^2 \gamma_x+ (\Lambda^2+\gamma_x^2) (\gamma_x-\gamma_y){\Big)}}{\Delta^4+(\Lambda^2+\gamma_x^2)(\Lambda^2+(\gamma_x-\gamma_y)^2)-2\Delta^2 (\Lambda^2+ \gamma_x(\gamma_y-\gamma_x))},
\end{align}

\begin{figure}
\includegraphics[scale=0.85]{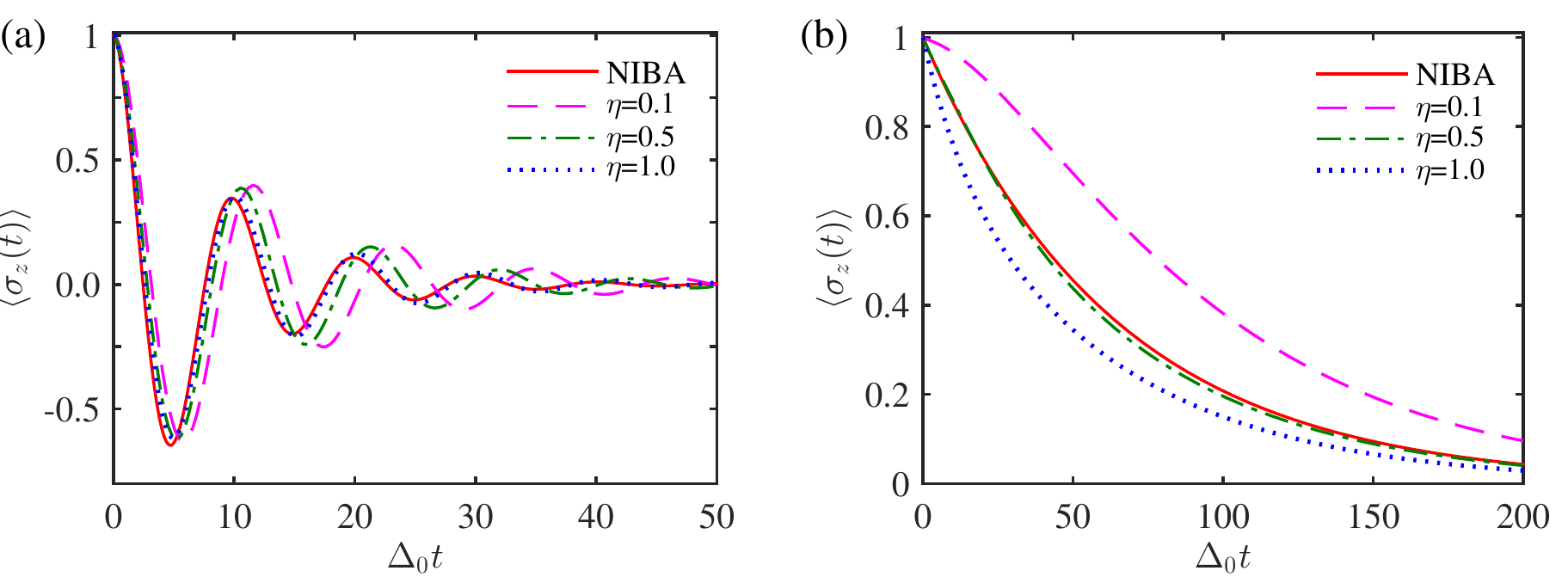}
\caption{\label{fig:eta}  (Color online) Comparison of TLS dynamics for three values of $\eta=\Lambda t$ at (a) $\alpha=0.10$  and (b) $\alpha=0.5$ (the Toulouse point). Here, $\Delta_0/ \omega_c= 0.01$.}
\end{figure} 

Next, we present the calculations for $\sigma_x$, starting from Eq.(\ref{eq:sigmaxRG1}). Let us first show that the upper line can be treated in a similar fashion to the above derivation and it ultimately yields the term $-\gamma (t) \langle\sigma_x(t)\rangle$.  Following the same steps, first, plug in the first order solution of $\sigma_y$, that is
\begin{align}
\label{eq:sigmaypert}
\sigma_y(t) =&  e^{-\gamma_y t/2} {\Big[} \sigma_y(0) {\big(} \text{cos}(\Omega t) - \frac{\gamma_y}{2 \Omega } \text{sin}(\Omega t) {\big)} +\sigma_z(0) \frac{\Delta}{\Omega } \text{sin}(\Omega t) {\Big]} \nonumber \\
&+ \sum_k g_k \int_0^t \text{d}t' e^{-\gamma_y (t-t')/2} {\Big(} \text{cos}(\Omega (t-t')) - \frac{\gamma_y}{2\Omega } \text{sin}(\Omega (t-t')) {\Big)}  (b_k(0)e^{-i\omega_k t'}+ b_k^\dagger(0)e^{-i\omega_k t'}) 
\sigma_x (t') + O(g_k^2);
\end{align}
Second, take the expectation value over bath modes, we get
\begin{align}
& \sum_k g_k^2 \ (2 n(\omega_k)+1) \int_0^t \text{d}t' e^{-\gamma_y (t-t')/2}  \text{cos}(\omega_k(t-t')) \ {\Big(} \text{cos}(\Omega(t-t')) - \frac{\gamma_y}{2\Omega} \text{sin}(\Omega(t-t')) {\Big)} \sigma_x(t');
\end{align}
Third, substitute the zeroth order solution of $\sigma_x$, i.e., $\sigma_x(t')=e^{-\gamma_x(t'-t)} \sigma_x(t)$. The expression can then be written in the form of ${\cal F}^1_x \langle \sigma_x(t)\rangle$, where
\begin{align}
{\cal F}^1_x= \sum_k g_k^2 \ (2 n(\omega_k)+1) \int^t \text{d}t' e^{-(\gamma_y -2\gamma_x) (t-t')/2}
\text{cos}(\omega_k(t-t'))\ {\Big(} \text{cos}(\Omega(t-t')) - \frac{\gamma_y}{2A} \text{sin}(\Omega(t-t')) {\Big)}.
\end{align}
The result of this integral is the same as ${\cal F}_y$ in Eq.~(\ref{eq:inty}). Hence, we conclude that $\gamma_x=\gamma_y=\gamma$. This should have been clear also from the beginning and consistent with the original equations of motion (Eqs (\ref{eq:sigmaxdot})-(\ref{eq:sigmazdot})) as there is no difference in the way $\sigma_x$ and $\sigma_y$ are coupled to the bath. Using this identity, now we can simplify the flow equations further into
\begin{align}
\label{eq:deltaRG} \frac{d\Delta}{d\Lambda} &= \frac{\Delta}{2\pi}\frac{J(\Lambda) {\Big(}\Delta^2-\Lambda^2{\Big)}}{(\Delta^2-\Lambda^2)^2+\gamma^2 \Lambda^2}, \\
\label{eq:gammaRG}\frac{d\gamma}{d\Lambda} &= \frac{\Delta^2}{\pi} \frac{\gamma J(\Lambda)  }{(\Delta^2-\Lambda^2)^2+\gamma^2 \Lambda^2}.
\end{align}

One should notice that the relation between time rescaling and frequency rescaling in DRG is logarithmic, i.e.~$dt/t=-d\Lambda/d\Lambda$. Therefore, the real-time is related to the frequency scale by $t=\eta/\Lambda$, where $\eta$ is a constant. Figure~\ref{fig:eta} shows two examples of how dynamics is affected as $\eta$ is changed.

Let us now focus on the second line of Eq.~(\ref{eq:sigmaxRG1}). After substituting the zeroth order solution of $\sigma_z$, that is 
\begin{align}
\sigma_{z}(t') =& e^{-\gamma_y (t'-t)/2} {\Big[} \sigma_z(t) {\big(} \text{cos}(\Omega(t'-t)) + \frac{\gamma_y}{2\Omega} \text{sin}(\Omega(t'-t)) {\big)} - \sigma_y(t) \frac{\Delta}{\Omega} \text{sin}(\Omega(t'-t)) {\Big]} + O(g_k), 
\end{align}
in the second line of Eq.~(\ref{eq:sigmayRG1}), it will read as
\begin{align} \label{eq:intx2}
{\cal F}_x^2=&\frac{1}{2} \sum_{k \in d\Lambda} g_k^2 \int^t \text{d}t' \text{sin}(\omega_k(t-t')) \ (\sigma_z(t')\sigma_y(t) + \sigma_y(t)\sigma_z(t')) \nonumber \\
=& \frac{1}{2} \sum_{k \in d\Lambda} g_k^2 \int^t \text{d}t' \text{sin}(\omega_k(t-t')) \ 
 e^{-\gamma_y (t'-t)/2} {\Big[} \{\sigma_z(t),\sigma_y(t)\} {\big(} \text{cos}(\Omega(t'-t)) - \frac{\gamma_y}{2\Omega} \text{sin}(\Omega(t'-t)) {\big)} +\{\sigma_y(t),\sigma_y(t)\} \frac{\Delta}{\Omega} \text{sin}(\Omega(t'-t)) {\Big]} \nonumber \\
 =& \mathbb{I}\ \frac{\Delta}{\Omega} \sum_{k \in d\Lambda} g_k^2 \int^t \text{d}t' \text{sin}(\omega_k(t-t')) \ 
 e^{-\gamma_y (t'-t)/2}  \text{sin}(\Omega(t'-t))  
\end{align}
where we use the anti-commuting property of the Pauli matrices (for equal-time TLS operators) $\{\sigma_i,\sigma_j\}=2\delta_{ij} \mathbb{I}$. Therefore, this term leads to the term ${\cal F}_x  \langle\mathbb{I}\rangle={\cal F}_x$, where
\begin{align}
{\cal F}^2_x=  \Delta \sum_{k\in d\Lambda} g_k^2 \frac{\gamma \Delta  }{(\Delta^2-\omega_k^2)^2+\gamma^2 \omega_k^2}.
\end{align}
Assuming that the equation of motion has the form of $\partial_t\langle \sigma_x\rangle=- \gamma\langle\sigma_x\rangle + \beta $, performing the above RG step  leads to the correction $\beta \to \beta + d\beta$ where the flow equation is given by  
\begin{align}
\frac{d\beta}{d\Lambda}= \frac{\Delta^2}{\pi} \frac{\gamma J(\Lambda)  }{(\Delta^2-\Lambda^2)^2+\gamma^2 \Lambda^2}.
\end{align}
Thus, we arrive at the same flow as $\gamma$ in Eq.~(\ref{eq:gammaRG}), which means $\beta=\gamma$.

The final remark is about the relation between our flow equations and those of the equilibrium RG. To arrive at the standard result, let us consider the usual form of the spectral function $J(\Lambda)=2\pi \alpha \Lambda \ \theta(\omega_c-\Lambda)\theta(-\Lambda)$, where a sharp cutoff $\omega_c\gg \Delta_0$ is introduced and neglect the decay term in the denominator. Assuming $\Delta \ll \Lambda$, Eq.~(\ref{eq:deltaRG}) can be simplified into
\begin{align}
\frac{d{\Delta}}{d \ln \Lambda}&=  -\alpha {\Delta} \nonumber \\
\Rightarrow \ln{\frac{{\Delta}}{\Delta_0}}&= \alpha \ln{\frac{\Lambda}{\omega_c}}  \nonumber \\ \label{eq:flow}
\Rightarrow {\Delta}(\Lambda)&= \Delta_0 \left( \frac{\Lambda}{\omega_c} \right) ^\alpha
\end{align}
The flow equation reaches the Kondo scale fixed point when $T_K\equiv \Delta = \Lambda $ and hence we have
\begin{align}
T_K &= \Delta_0 \left( \frac{T_K}{\omega_c} \right) ^\alpha \nonumber \\
\Rightarrow T_K &= \Delta_0 \left( \frac{\Delta_0}{\omega_c} \right)^{\frac{\alpha}{1-\alpha}}
\end{align}
that is the well-known Kondo scale for the ohmic spin-boson model~\cite{Leggett1987}. This gives us a time-scale at which the TLS parameters approach the long-time final values.


\section{\label{sec:SE} The spontaneous emission problem}

In this appendix, we show the DRG results for the spontaneous emission problem which can be solved exactly.
In quantum optics literature, it is customary to make the rotating-wave approximation and drop the terms of type $b_k \sigma_-$ and $b_k^\dagger \sigma_+$~\cite{Scully2001}. Thus, the Hamiltonian becomes
\begin{equation} \label{eq:Hspon}
H_{SE}= \frac{\Delta_0}{2} \sigma_z + \sum_k \omega_k b_k b_k^\dagger + \frac{1}{2} \sum_k g_k (\sigma_+ b_k+b_k^\dagger\sigma_-)
\end{equation}
which is also known as the celebrated Wigner-Weisskopf theory of spontaneous emission. Note that in going from the basis introduced for the spin-boson model in Eq.~(\ref{eq:SBmodel}) to the usual form of the spontaneous emission problem above, we need to make a spin rotation around the y-axis such that $\sigma_x\to \sigma_z$ and $\sigma_z\to -\sigma_x$. The approximation here is justified when $\alpha  \lesssim \Delta_0/\omega_c$. It turned out that in this case one can write an exact master equation in the Lindblad form for TLS density operator ${\rho}_s$~\cite{Breuer2007,Hu2000}
\begin{equation} \label{eq:Mexact}
\frac{\partial{\rho}_s}{\partial t} = -i \Delta(t) [\sigma_z, {\rho}_s] + \gamma(t) {\cal L}[{\rho}_s],
\end{equation}
where ${\cal L}[\rho]=2\sigma_-\rho\sigma_+ - \sigma_+\sigma_-\rho - \rho\sigma_+\sigma_-$ is the Liouvillian super-operator. The time-dependent TLS frequency $\Delta(t)$ and decay rate $\gamma(t)$ are given by the expression 
\begin{align}
\frac{\dot{u}(t)}{u(t)}= - \gamma(t)- i\Delta(t),
\end{align}
 where $u(t)$ is a solution of the integrodifferential equation 
\begin{align}
\dot{u}(t) + i \Delta_0 u(t)+ \frac{1}{2} \int_0^t \text{d}t' \mu(t-t') u(t')=0.
\end{align}
The kernel function $\mu(\tau)=\sum_k g_k^2 \text{exp}(-i\omega_k \tau)$ is the inverse Fourier transform of the spectral function in Eq.~(\ref{eq:spec}). This is a quite important remark that TLS dynamics can be exactly determined through a Born-Markov form solution with time-dependent parameters; indeed, our method is originally motivated by this observation. 
Therefore, the dynamics of TLS is governed by the following equations
\begin{align}
\partial_t \langle{\sigma}_\pm \rangle&= (\pm i \Delta(t) -\gamma(t)) \langle\sigma_\pm \rangle,\\
 \partial_t \langle{\sigma}_z \rangle&=- 2\gamma(t) (\langle\sigma_z\rangle+1).
\end{align}
Following the steps in Appendix~\ref{sec:DRG_der}, one can obtain the flow equations as
\begin{align}
\frac{d\Delta}{d\Lambda}&=  \frac{J(\Lambda)}{4\pi} \frac{(\Delta-\Lambda)}{(\Delta-\Lambda)^2+\gamma^2} \\
\frac{d\gamma}{d\Lambda}&= \frac{J(\Lambda)}{4\pi} \frac{\gamma}{(\Delta-\Lambda)^2+\gamma^2} .
\end{align}
Figure~\ref{fig:compSE} shows a comparison between the DRG method and the exact solution.  It is interesting to note that in Fig.~\ref{fig:compSE}(c) the Born-Markov approximation agrees with the exact results only for small 
$\alpha$ and looks more like a linear approximation to the exact results.

\begin{figure}
\includegraphics[scale=0.45]{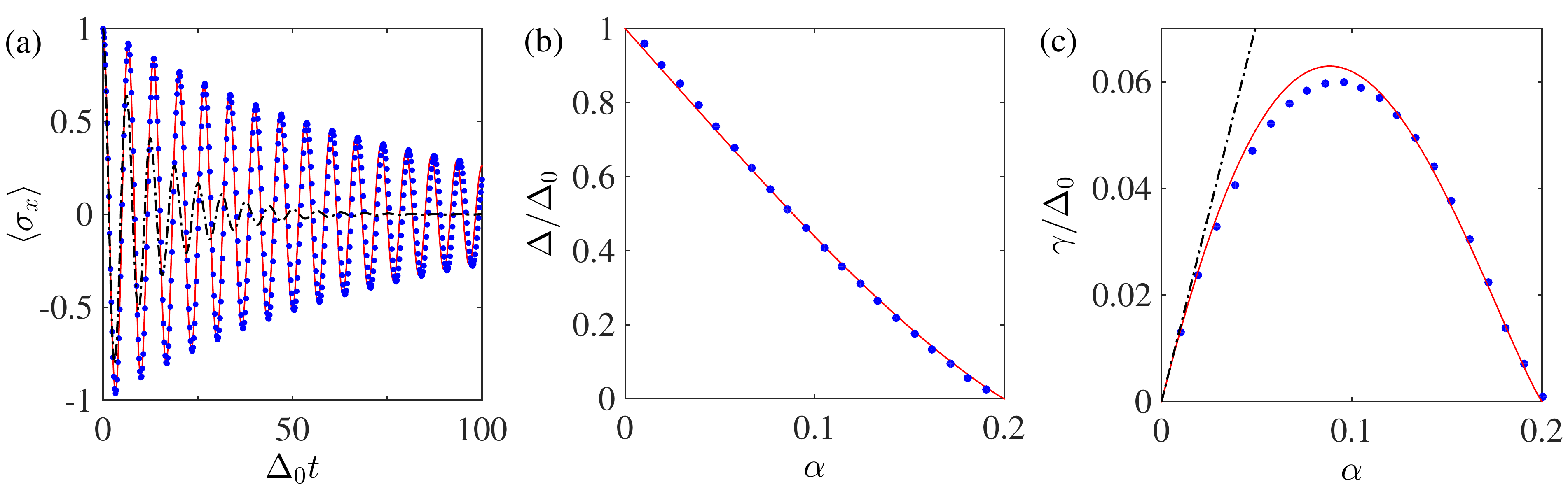}
\caption{\label{fig:compSE} (Color online) (a) Dynamics of TLS in the spontaneous-emission problem, Eq.~(\ref{eq:Hspon}), for $\alpha=0.05$ and $\Delta_0/\omega_c=0.1$. The long-time renormalized TLS  (b) frequency and (c) decay rate as a function of coupling constant $\alpha$.
In all three graphs, DRG is blue circles, the exact solution is solid red and the Born-Markov approximation is dash-dotted black. }
\end{figure}

\end{widetext}
\end{appendix}

\bibliography{DRG_v8}

\end{document}